\documentclass[a4paper,11pt,twoside]{llncs}
\usepackage{graphicx}				
\usepackage{amsmath}				
\usepackage{amsfonts}
\usepackage{url}				
\usepackage[colorlinks=true,linkcolor=black,anchorcolor=black,citecolor=black,filecolor=black,menucolor=black,urlcolor=black,breaklinks=true,pdfhighlight=/P,pdfmenubar=true,pdftoolbar=true,pdfpagelabels=true,pdfstartpage=1,pdfstartview=FitV,pdftitle={Titel},pdfsubject={Thema},pdfauthor={Autor},pdfcreator={Ersteller},pdfproducer={Produzent},pdfkeywords={Stichworte}]{hyperref}	
\usepackage[numbers,sort&compress]{natbib}	
\usepackage{hypernat}				
\usepackage[american]{babel}		
\begin{document}
%
\selectlanguage{american}
\title{The use of dynamic distance potential fields for pedestrian flow around corners }
\toctitle{The use of dynamic distance potential fields for pedestrian flow around corners }
\author{Tobias Kretz}
\institute{PTV AG, Stumpfstra{\ss}e 1, D-76131 Karlsruhe\\\email{Tobias.Kretz@PTV.De}}

\maketitle


\begin{abstract}
This contribution investigates situations in pedestrian dynamics, where trying to walk the shortest path leads to largely different results than trying to walk the quickest path. A heuristic one-shot method to model the influence of the will to walk the quickest path is introduced.
\end{abstract}

\section{Introduction}
In 1952 John Wardrop raised his first principle that ``The journey times in all routes actually used are equal and less than those which would be experienced by a single vehicle on any unused route."\cite{Wardrop1952} which since has become a foundation of transport planning. But while all kinds of static and dynamic assignment methods \cite{Leonard1989,Mahmassani1995,Benakiva1997,Bargera2003,VISSIM2008} have been invented to calculate the user equilibrium of least journey time on road networks, in models of pedestrian dynamics \cite{Schadschneider2009} journey time mostly matters only indirectly. Typically the influence of following the shortest path is modeled explicitly and the influence of the quickest path only indirectly with costs, reduced probabilities, or repelling forces, if an agent comes too close to another one, a group of pedestrians or an obstacle. These implicit methods may have some effect for pushing agents more toward the quickest path. Additionally there are reasons, why journey time might matter less to pedestrians than for vehicle drivers, as attractiveness or safety of a route or just having plenty of time wanting to avoid the extra physical effort needed to walk on the quickest path may play a (more important) role. But considering the vast discussion of journey times in vehicular traffic, it appears that a discussion of the explicit influence of the will to minimize the expected remaining journey time is underrepresented in 2d models of pedestrian dynamics. Nevertheless, there is discussion on considering travel time and user equilibrium for network \cite{Haupt2006,Laemmel2008} and macroscopic \cite{Hughes2002,Xia2008} models of pedestrian flow.

For the simulation of pedestrian dynamics things get more complicated, as basically there are two methods to consider journey time: either fully two dimensional without space semantic, respectively with full semantic symmetry between all accessible coordinates (this is discussed in \cite{Hoogendoorn2004b}) or with a semantic decomposition and assignment of coordinates to special functional and semantic structures of the geometry i.e. the calculation of a meaningful graph, which reduces the details of reality for the planning process. One may assume that when planning a path humans probably use something in between those alternatives: a two-dimensional representation of the space directly visible to them and a graph representation of anything that's out of view.

Generally, if the journey time itself is meant to be a determining factor for agents' behavior, there are only two methods of solution: analytically or iterative. There can not be a one-shot simulation considering journey time as input, as only at the end of the simulation the journey time is known. An analytic solution obviously is out of reach for any real-life application. An iterative solution is in principle possible, but in fact often impossible, if one considers typical computation times of simulations of pedestrian and evacuation dynamics. This leaves one-shot simulations using not journey time, but heuristics, and methods using simplifying assumptions like static assignment.

It is obvious that the assumptions made for static assignment do not hold for evacuations. During evacuations there is not a constant, but a strongly time-dependent, short but strong demand, almost delta-function-like, widened by varying awareness times. During the egress time period of one individual the best decision -- in terms of journey time -- where to turn to at some decision spot, might change a number of times. The situation may be compared to the one of rush hours in vehicular traffic, and for these dynamic assignment methods were invented at first. 

Following this line of hand-waving argumentation, a heuristic semanticless method to model effects of the will to move on the path of least expected remaining journey time has been introduced \cite{Kretz2009} and will be sketched and investigated here. With reliance on a heuristic rather than estimated remaining travel time itself, in a strict sense the initial demand to consider the estimated travel time for walking behavior is given up. There are two arguments that may justify this: ``estimation" in reality brings in a somewhat fuzzy element, such that real pedestrians also do not rely on a precise value of remaining travel time. It is not guaranteed, but it might even be that the heuristic discussed in this contribution might come closer to reality than a costly calculated remaining travel time. The second argument is that the results justify that it's more realistic to use the method than not to use it and if it's computationally faster than any method yielding the same benefit.


\section{Description of the Method}
Let's take a look at a group of pedestrians forming a jam at some bottleneck, where alternative paths to detour around this group are still available for other pedestrians approaching from behind. Assume the extra effort of walking the detour is reasonably small, such that a considerable ratio of real pedestrians would choose to walk it. How can it be modeled efficiently that agents of a simulation reproduce this behavior?

In principle potentials \cite{Kimmel1998,Nishinari2004,Kretz2008c} should do the task by just taking jams into accounts like obstacles. But problems arise with details: 
\begin{enumerate}
\item Some methods for potential calculation are not fast enough execution each time step.
\item One has to deal with the case, when no detour is available, i.e. the jam would block the potential from spreading to some regions.
\item The potential calculation method must be suited to consider the size of a jam quantitatively. A single agent must have only a negligible effect, a large jam should have a wide spreading effect.
\end{enumerate}

All this can be achieved using a flood fill, where a value of 1 is added, if a cell is unoccupied and some larger value $s_{add}$, if it is occupied by an agent. But the two simple flood fill methods result in Manhattan or Chebyshev but not the desired Euclidean metric. In this sense these two methods produce large errors and unwanted artifacts in the movement. However, the errors can be reduced for once by combining the two methods and second by using as heuristic not the time dependent potential, but its difference to the potential calculated on the empty (unoccupied) geometry.

For two mutually visible coordinates $(x_0,y_0)$ and $(x_1,y_1)$, Manhattan distance is
\begin{equation}
d^M=|x_1-x_0|+|y_1-y_0|=|\Delta x| + |\Delta y|
\end{equation}
and Chebyshev distance is
\begin{equation}
d^C=\max(|\Delta x|,|\Delta y|)
\end{equation}
From this the ``minimum distance" follows:
\begin{equation}
d^m=d^M-d^C=\min(|\Delta x|,|\Delta y|)
\end{equation}
and therefore the Euclidean distance is
\begin{equation}
d^E=\sqrt{(d^M)^2+(d^m)^2}=\sqrt{|\Delta x|^2+|\Delta y|^2}
\end{equation}
Combining the whole potentials cell by cell in this manner -- even if cells are not mutually visible -- will now be called ``method V1". A discussion of the resulting errors of this method compared to Euclidean distance is given in \cite{Kretz2008c}.

Let's call the potential calculated on the empty (no agents) geometry $S_{V1}^0$ and the potential after a certain time step $S_{V1}(t)$. Then the heuristics influencing the motion of pedestrians is $S_{dyn}(t)=S_{V1}(t)-S_{V1}^0$.

This has been used and coupled to the F.A.S.T. model \cite{Kretz2006f,Kretz2006d,Kretz2006k,Kretz2007a,Kretz2008f} as an additional partial probability $p_{dyn}=e^{-k_{Sdyn}S_{dyn}(t)}$, multiplied to the original probability and normalized accordingly.

\section{Example of the Method's Effect}
\begin{figure}[htbp]
  \center
	\includegraphics[width=0.62\textwidth]{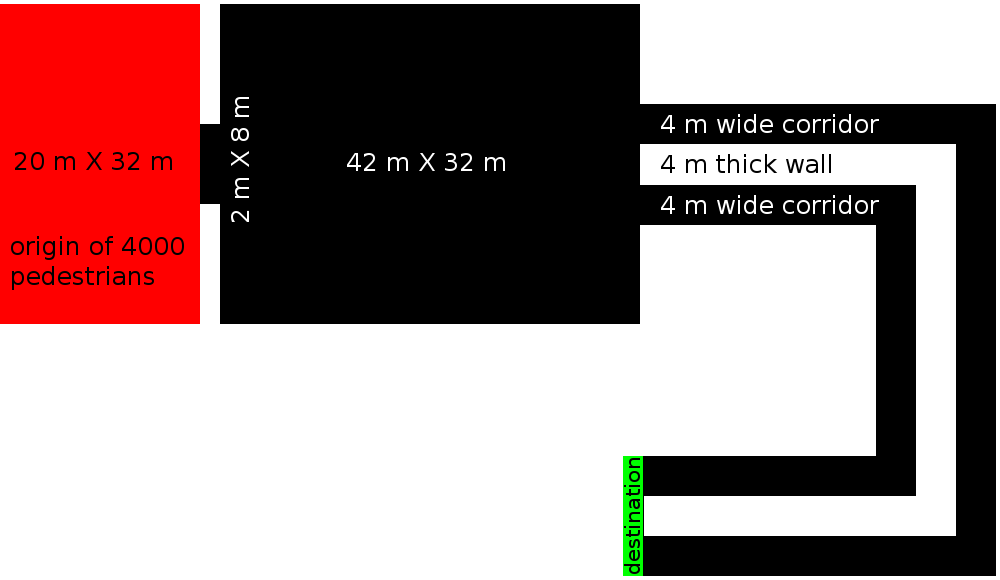}
	\caption{Agents have to walk from the red to the green area, thereby choosing between one of the two corridors, of which one is 32 m longer.}
	\label{fig:scenario}
\end{figure}

Figure \ref{fig:scenario} shows the scenario that is used in this contribution to investigate the effect of the method. 4000 agents have to walk from the red to the green area. They have to choose one of the two corridors. The longer corridor is 32 m longer than the short one, which at the median maximum speed of the agents of $v_m=1.6 m/s$ equals to an additional 20 seconds. Using the shorter corridor, the shortest total distance for an agent to walk is about 325 m, the longest about 418 m. This leads to a relative additional distance of 7.6\% to 9.8\%.

\begin{figure}[htbp]
  \center
	\includegraphics[width=0.62\textwidth]{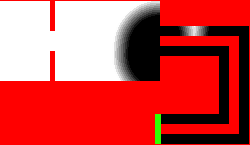}
	\caption{Static potential with contrast and brightness adjusted to show the point inside the corridors, where the destination is equally far away, no matter, if an agent walks to the left or right.}
	\label{fig:staticpotential}
\end{figure}

Figure \ref{fig:staticpotential} shows that there is no chance that without using a dynamic distance potential field an agent will move over the local maximum of distance toward the destination (i.e. local maximum of the static floor field) inside the longer corridor. Thus, in this case all agents will use the shorter corridor.

\begin{figure}[htbp]
  \center
	\includegraphics[width=0.62\textwidth]{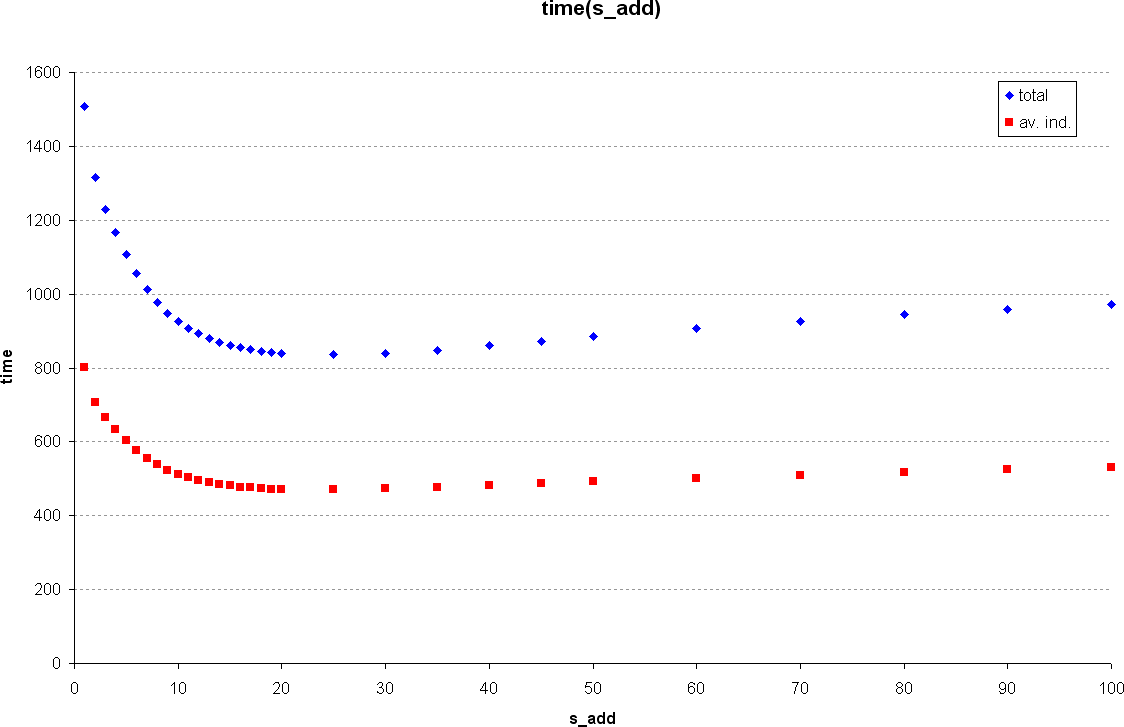}
	\caption{Total times and average individual egress times in dependence of parameter $s_{add}$. For each value of $s_{add}$ 100 simulations were carried out and the average was calculated. With increasing $s_{add}$ the total time and the average individual egress time slowly keep increasing. $k_{Sdyn}$ has been set to $k_{Sdyn}=1.0$ for these simulations.}
	\label{fig:results1}
\end{figure}

\begin{figure}[htbp]
  \center
	\includegraphics[width=0.45\textwidth]{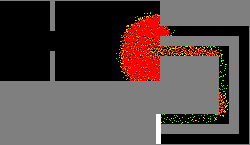} \hspace{10pt}
	\includegraphics[width=0.45\textwidth]{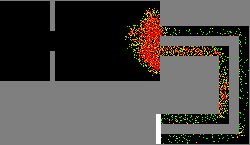}
	\caption{Situation after 600 seconds without dynamic distance potential field (left) and with $k_{Sdyn}=1.0$ and $s_{add}=10$ (right).}
	\label{fig:spatial}
\end{figure}

For this scenario the method would cause, what is intended, if the load of the two corridors is shifted toward an equilibrium, implying a reduced total time for the process as well as a reduced average individual egress time of the agents. Figure \ref{fig:results1} shows the values of these two observables in dependence of $s_{add}$ and the strong influence the method has, and figure \ref{fig:spatial} directly shows the effect on route choice behavior.

\begin{figure}[htbp]
  \center
	\includegraphics[width=0.62\textwidth]{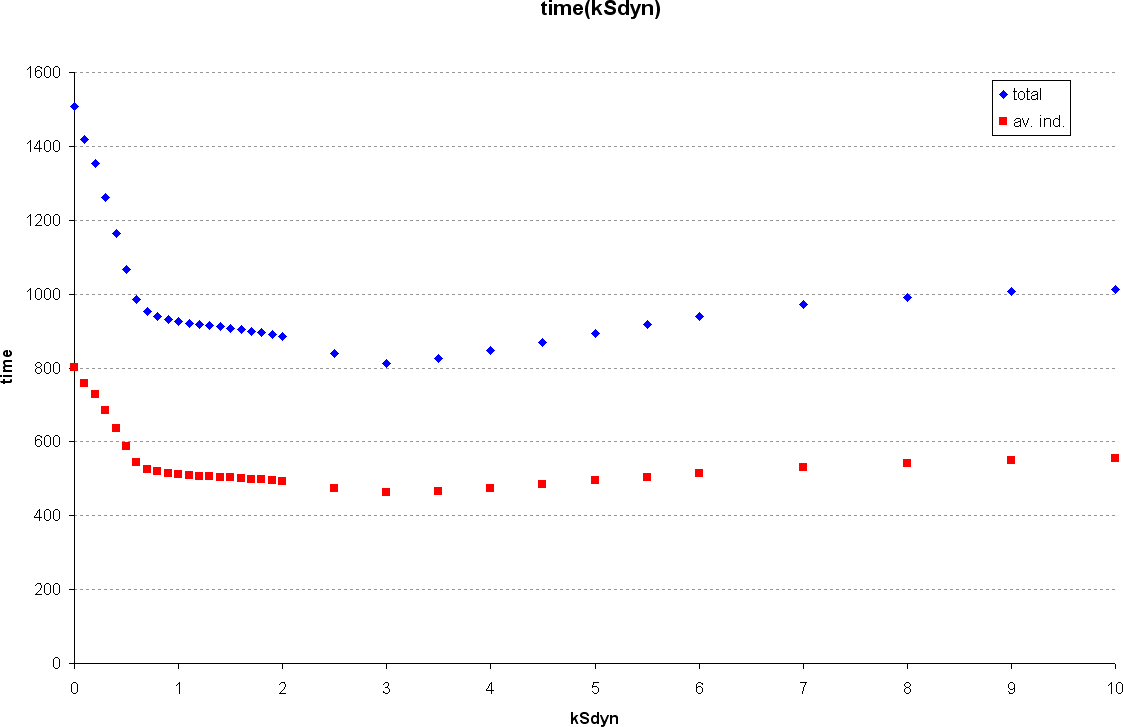}
	\caption{Total times and average individual egress times in dependence of parameter $k_{Sdyn}$. Each value of $k_{Sdyn}$ is an average of 100 simulations. For larger values of $k_{Sdyn}$ the total time approaches 1020 seconds. $s_{add}$ has been set to $s_{add}=10$ for these simulations.}
	\label{fig:results2}
\end{figure}

It is interesting to note that for $s_{add}=2$ and $l_{Sdyn}=1.0$ not a single agent walks the longer corridor, still the two measured times drop significantly by about 12\%. This is because the agents moved around the corners inside the corridor more efficiently. For higher values of $s_{add}$ more and more agents walk the longer corridor. This means that two time reducing effects are active. 

The dependence on parameter $k_{Sdyn}$ surprisingly showed a more complicated behavior, which is shown in figure \ref{fig:results2}. Between $k_{Sdyn}=1.0$ and $k_{Sdyn}=2.0$ the evacuation times seem to stabilize and settle, just to start decreasing again at slightly higher values, pass a minimum and increase again and finally stabilize for very large values of $k_{Sdyn}$.

\begin{figure}[htbp]
  \center
	\includegraphics[width=1.0\textwidth]{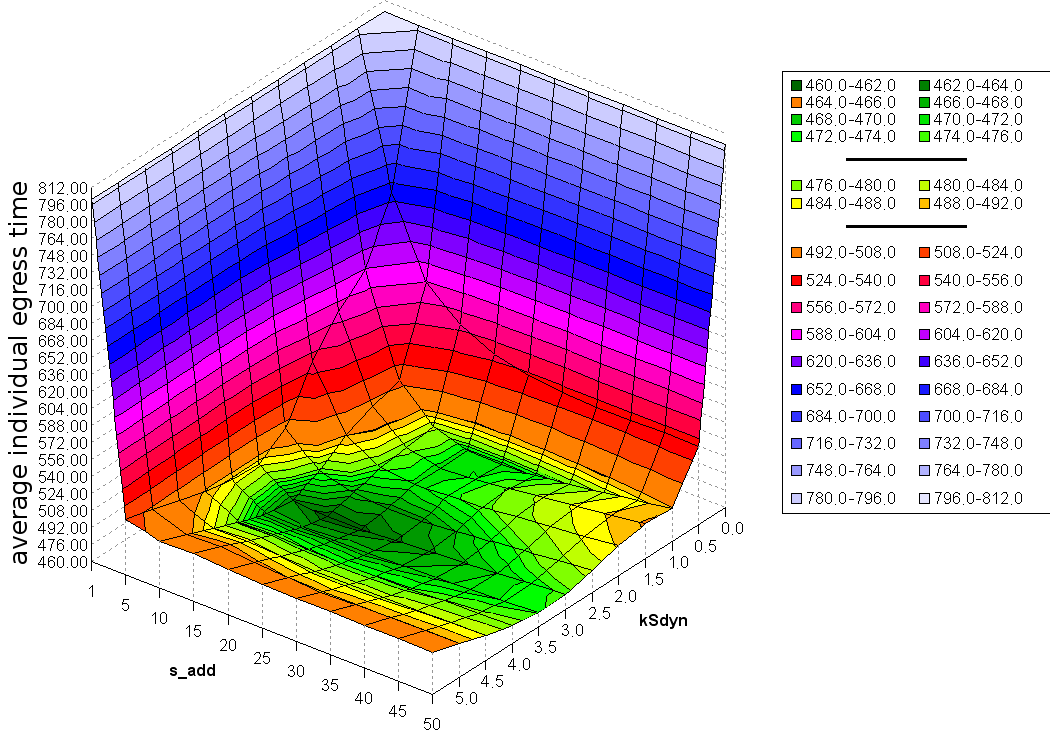}
	\caption{Average individual egress times in dependence of $k_{Sdyn}$ and $s_{add}$.}
	\label{fig:results3}
\end{figure}

A result as in figures \ref{fig:results1} and \ref{fig:results2} calls for a combined investigation of the two evacuation times' dependence on $k_{Sdyn}$ and $s_{add}$. The result of such an investigation for the average individual egress time is shown in figure \ref{fig:results3}. The result for the total times is very similar, but the standard deviations are larger, as the arrival time of the last few agents varies increasingly with increasing product $k_{Sdyn} \cdot s_{add}$ (the last few agents hesitate at the point of aequi-distance). 

In this two-dimensional parameter space the results look a bit complex as well. There are two local minima. (For larger values of the parameters, which cannot be displayed without obscuring the interesting region, the behavior qualitatively is the same as in figures \ref{fig:results1} and \ref{fig:results2}.)  

\begin{figure}[htbp]
  \center
	\includegraphics[width=1.0\textwidth]{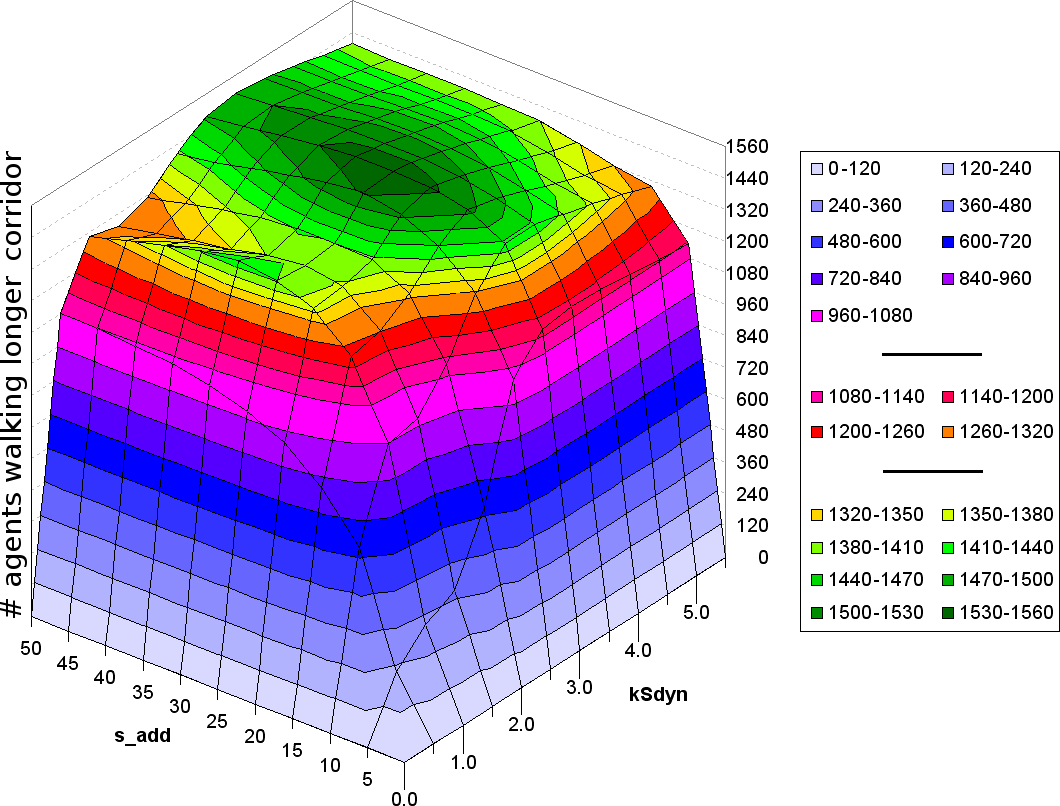}
	\caption{Average number of agents walking the longer corridor in dependence of $k_{Sdyn}$ and $s_{add}$. Please note that this plot is rotated by 180 degree compared to figure \ref{fig:results3}.}
	\label{fig:numbers}
\end{figure}

At latest at this point one is interested in the number of agents walking through the longer corridor and the dependence of this number on the two parameters controlling the method. These numbers are shown in figure \ref{fig:numbers}. With a maximum (at $s_{add}=25$ and $k_{Sdyn}=3.0$) of 1543.8 agents using the longer corridor, the method achieves to do a considerable step toward the user equilibrium (which might be above 1900 agents). 

The second observation is that this plot just as well has two extrema (maxima in this case). This implies that the two maxima in figure \ref{fig:results3} are not a result of the double influence of the method on local operational behavior (walking around a jam at a corner) and route choice behavior. 

\begin{figure}[htbp]
  \center
	\includegraphics[width=1.0\textwidth]{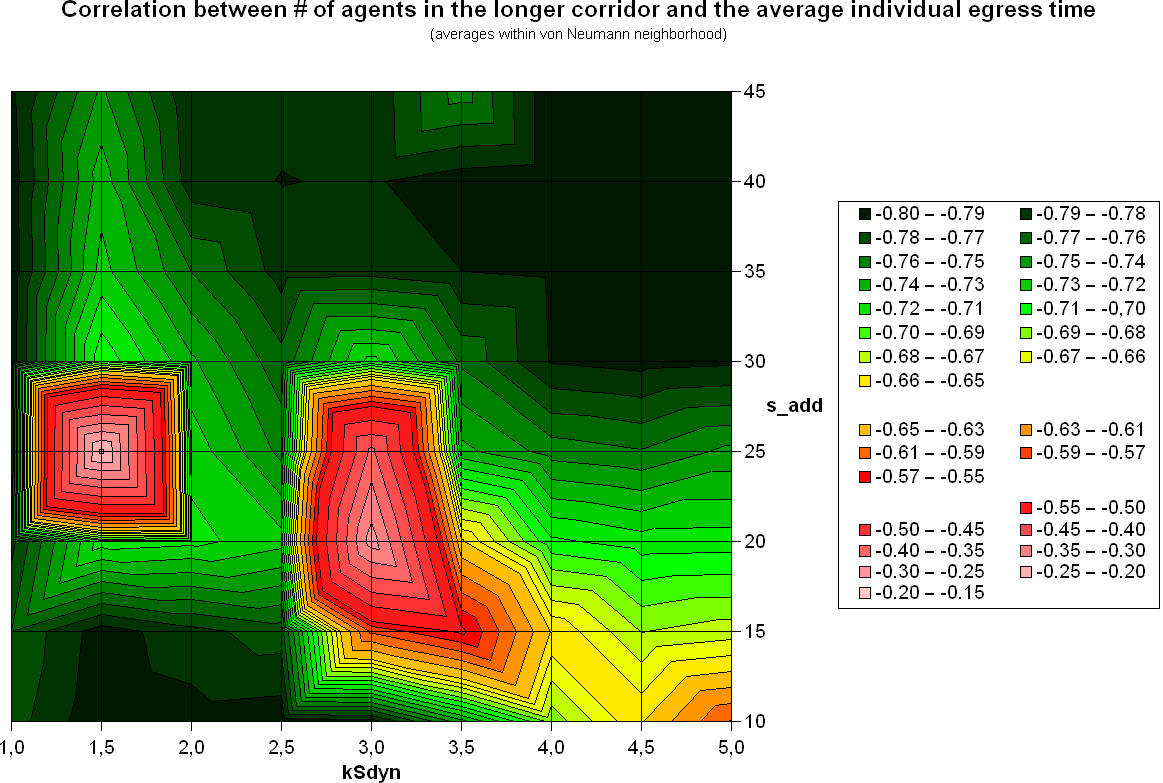}
	\caption{This figure shows the correlation between average individual egress time and load in the longer corridor of a data point and its nearest neighbor. Including the next to nearest neighbors (Moore neighborhood) does not change the result qualitatively, but makes it even more extreme quantitatively (the peak at $k_{Sdyn}=1.5$ and $s_{add}=25$ is actually positive).}
	\label{fig:correlations}
\end{figure}

On the other hand there's only a limited anti-correlation between the number of agents walking the longer corridor and the average individual egress time (see figure \ref{fig:correlations}). This is a hint that increasing $k_{Sdyn}$ at some point leads to artifacts in the motion of the agents, which on average reduce the progress speed and therefore more agents need to walk the longer corridor to give the same average individual egress time as at lower $k_{Sdyn}$. In table \ref{tab:compare1} pairs of results are compared, where either the numbers of agents walking the longer corridor are very similar and the average individual egress time is larger, where $k_{Sdyn}$ is larger or vice versa the average individual egress times are similar and the number of agents in the longer corridor is smaller for smaller $k_{Sdyn}$. Interestingly for identical $k_{Sdyn}$ and two different $s_{add}$, there is no such effect, i.e. either both measured values are very similar or none.

\begin{table}
\center
\begin{tabular}[htbp]{c|ccc|ccc}
$s_{add}$ & $k_{Sdyn}$ &  time & load   & $k_{Sdyn}$ & time  & load \\ \hline
    5     &    2.5     & 501.4 & 1044.5 &     4.5    & 498.1 & 1161.3 \\
    10    &    1.5     & 503.0 &  983.9 &     5.5    & 504.4 & 1297.3 \\
    10    &    2.5     & 474.9 & 1315.0 &     4.0    & 474.9 & 1404.9 \\
    15    &    2.0     & 475.3 & 1315.0 &     4.0    & 475.3 & 1438.9 \\
    25    &    1.5     & 472.3 & 1398.8 &     4.0    & 473.9 & 1487.7 \\
    50    &    1.0     & 491.6 & 1302.2 &     5.0    & 492.1 & 1423.5 \\ \hline
    25    &    1.5     & 472.3 & 1398.8 &     4.0    & 473.9 & 1487.7 \\
    35    &    2.0     & 473.4 & 1409.0 &     5.0    & 492.0 & 1412.5 \\ 
\end{tabular}
\caption{Comparison of similar results for different combinations of $k_{Sdyn}$ and $s_{add}$. ``time" is the average individual egress time and ``load" the number of agents passing the longer corridor.}
\label{tab:compare1}
\end{table}

\section{Summary}
A non-iterational method was introduced that intends to shift the operational and tactic route choice behavior of agents in a simulation of pedestrian dynamics from a strong focus on the shortest path toward user equilibrium and a consideration of the quickest path. The investigation with a proto-typic geometry gave promising results and showed that in general the idea works as originally thought off.

However, in detail the results were more complicated than expected as dependencies on parameters showed up to be non-monotonic, yielding up to three local extrema. The cause of this behavior could not be clarified entirely.

A technical result is that values for parameter $k_{Sdyn}$ that make agents' behavior look most natural do not realize the full potential of the method to come close to the user equilibrium. In turn this means that with parameters lead to a route choice as close as possible to the user equilibrium, the operational behavior includes unrealistic artifacts from the construction of the dynamic distance potential field.

It's in the nature of the method that the last agent leaving the longer corridor will always do this earlier than the last agent leaving the shorter corridor. But having these two agents doing this almost simultaneously is one criterion for Wardorp's first principle being fulfilled. I.e. with this method the last agent passing the shorter corridor would always have arrived earlier, if he had joined the last agent passing the longer corridor. Like already stated above: Wardrop's first principle is one thing, the question, how real pedestrians would behave, is different one. But if one takes the theoretical point of view, eager to have a simulation that leads to a result, where Wardrop's first principle is fulfilled, the method proposed in this contribution is a good starting point for an iterated dynamic assignment search for an equilibrium. It will -- in a computationally efficient way -- reduce the number of iterations to find the equilibrium. 

\section{Acknowledgments}
The author is grateful to Peter Vortisch for valuable hints and discussions on assignment issues.
%
\nocite{_PED2005,_ACRI2006,_ACRI2008,_PED2008}
\bibliographystyle{utphys_quotecomma}
\bibliography{Kretz_ICEM09}
%
\end{document}